\documentclass{article}

\usepackage[final]{neurips_2019}
\usepackage[utf8]{inputenc} 
\usepackage[T1]{fontenc}    
\usepackage{hyperref}
\usepackage{url}            
\usepackage{booktabs}       
\usepackage{amsfonts}       
\usepackage{nicefrac}       
\usepackage{microtype}      
\usepackage{mathtools}
\usepackage{hhline}
\usepackage{xcolor}

\newcommand{\xhdr}[1]{\vspace{0.2mm}\noindent{{\bf #1.}}}

\title{An Empirical Study on Learning Fairness Metrics for COMPAS Data with Human Supervision}

\author{
  Hanchen Wang\\
  Department of Engineering\\
  University of Cambridge\\
  \texttt{hw501@cam.ac.uk} \\
  \And
  Nina Grgi\'{c}-Hla\v{c}a\\ 
  Max Planck Institute for\\
  Software Systems\\
  \texttt{nghlaca@mpi-sws.org}
  \And
  Preethi Lahoti\\
  Max Planck Institute \\
  for Informatics\\
  \texttt{plahoti@mpi-inf.mpg.de}
  \And
  Krishna P. Gummadi\\
  Max Planck Institute for\\
  Software Systems\\
  \texttt{gummadi@mpi-sws.org}
  \And
  Adrian Weller\\
  University of Cambridge \\
  \& The Alan Turing Institute \\
  \texttt{aw665@cam.ac.uk} \\
}

\begin{document}

\maketitle
\begin{abstract}
  The notion of individual fairness requires that similar people receive similar treatment. However, this is hard to achieve in practice since it is difficult to specify the appropriate similarity metric. 
  In this work, we attempt to learn such similarity  metrics from human annotated data. We gather a new dataset of human judgments on a criminal recidivism prediction (COMPAS) task.
  Assuming that people's judgments encode the fairness metric they adhere to, we leverage prior work on metric learning and attempt to learn people's similarity metrics from these judgments.
\end{abstract}

\section{Introduction}
Bias in automated decision making systems has raised many concerns. One approach to address these concerns is to enforce individual fairness \citep{individualfairness}, which requires treating similar people similarly. However, it is not straightforward to quantify the appropriate similarity of individuals. There have been some noteworthy subsequent works on this topic, such as \citep{zemel2013learning} and \citep{lahoti2019ifair}. In our paper, we study the problem of learning an individual fairness metric from human annotated data. We leverage the intuition that human judgments might implicitly encode an underlying fairness metric they adhere to. We gather a dataset of human judgments about criminal recidivism risk predictions, and utilize this dataset to test the performance of several different metric learning algorithms on this data.

\section{Related Work}

\xhdr{Algorithmic Fairness}
Most prior work on algorithmic fairness has focused on group notions of fairness, which require that protected groups as indicated by sensitive attributes receive similar treatment to others \citep{zafar_fairness,zafar_dmt,hardt2016equality}. On the other hand, individual fairness, introduced by \citet{individualfairness}, requires that similar individuals be treated similarly. To achieve this, one must first define a similarity metric that can be used to compare the individuals. This similarity metric needs to be either given or learned from data. Some recent work on individual fairness \citep{lahoti2019operationalizing,jung2019eliciting} elicit human judgments for individual fairness, and incorporate them as pairwise constraints in the learning objective, without explicitly learning a fair distance metric.
Other works like \citep{speicher2018unified,kearns2017meritocratic,liu2017calibrated} use the objective of the learning algorithm itself as an implicit similarity metric in the optimization problem.
Instead of specifying a similarity metric directly, we attempt to learn it from human annotated data.

\xhdr{Metric Learning}
We leverage the rich literature on metric learning, which has been studied and applied in various domains, ranging from image processing \citep{fei2005bayesian} to recommendation systems \citep{mcfee2012learning}. In the past years, it has increasingly been applied on human annotated data, in order to model human notions of similarity \citep{tamuz2011adaptively}. In our work, we also take this approach.
We refer the reader to \cite{bellet2013survey} and \cite{kulis2013metric} for a more in depth overview of the relevant literature on metric learning.

\xhdr{Learning Fairness Metrics}
The recent work of \citet{ilvento2019metric} is closest to ours. They propose an approach for approximating an individual fairness metric from human judgments about the relative distance between inputs. While prior work did suggest that humans find it easier to make relative judgments than absolute ones \citep{stewart2005absolute}, data on relative judgments might be more difficult to obtain than data on absolute judgments.
For instance, recommendation systems collect data about which articles and movies users click on, and not about the users' relative preferences between these articles or movies. Therefore when learning fairness metrics in real world applications, it may be easier to utilize existing data about people's past absolute judgments, than to gather new data about relative judgments. 
Hence, in our work, we study the possibility of learning fairness metrics from absolute ratings.

\section{Methodology}

\subsection{Gathering Human Judgments}
\xhdr{Scenario}
In our experiments, we focus on the task of predicting criminal recidivism risk. We use a dataset related to the COMPAS tool -- a tool used across the United States to help judges make bail decisions by predicting defendants' criminal recidivism risk on a 10 point scale \citep{propublica_story}. This dataset, gathered by ProPublica \citep{propublica_story}, contains information about the recidivism risk predicted by the COMPAS tool, as well as the ground truth recidivism rates, for 7214 defendants who were arrested in Broward County, Florida, in 2013 and 2014.

\xhdr{Survey Instrument}
To gather human judgments, we conducted an online survey in which we asked participants to estimate the likelihood of criminal recidivism of a fixed set of 200 defendants from the ProPublica dataset.\footnote{These 200 defendants were selected uniformly at random without replacement from the 1000 ProPublica cases which were studied by \cite{dressel2018accuracy}, in order to allow us to compare the performance of our respondents with the performance of their respondents, on the same set of cases.}
To mitigate the effects of order bias \citep{redmiles2017summary}, the defendants were shown in random order. For each defendant, participants were shown information about the defendant's demographics and criminal history, in the same format as by \cite{dressel2018accuracy} and \cite{grgic2019mahdm}, and were asked to answer three questions: (Q1) \emph{How likely do you think it is that this person will commit another crime within 2 years?}, (Q2) \emph{Do you think this person should be granted bail?}, and (Q3) \emph{How confident are you in your answer about granting this person bail?}. Our participants were asked to respond to question (Q2) with \emph{yes} or \emph{no}, and to questions (Q1) and (Q3) using a 5-point Likert scale. Even though the COMPAS tool provides criminal recidivism risk predictions on a 10-point scale, we opted for this design choice in order to minimize the duration of our 200-question survey, since providing answers using 10-point Likert scales was found to be more time consuming than using 5-point scales \citep{matell1972there}.

\xhdr{Procedure}
We recruited participants through the online crowdsourcing platform Prolific \citep{palan2018prolific}. Utilizing Prolific's advanced pre-screening options, we recruited 29 participants from the US who self-reported to have served on a jury. On average, the respondents took approximately 71 minutes to complete the survey, and were paid a base fee of \pounds8.50. 
In order to increase response quality \citep{vaughan2017making}, we also provided a performance-based bonus.\footnote{Performance-based payments have been found to increase the quality of responses in effort-responsive tasks \citep{vaughan2017making}. Hence, in order to incentivize participants to provide high-quality survey responses, we increased their bonus fee by \$0.10 for each correct bail decision (i.e., when their response to (Q2) coincided with the ground truth data about the defendant's two year recidivism), and decreased it by the same amount for each incorrect bail decision.}
As additional quality control measures, we discarded responses of participants who (i) did not respond to our 5 attention check questions correctly, or (ii) completed the survey in less than 45 minutes. After discarding the responses of these participants, our final sample consisted of 20 participants. These 20 participants had an average criminal recidivism prediction accuracy of 62.4\%, close to the 62.1\% and 60.2\% that \cite{dressel2018accuracy} and \cite{grgic2019mahdm} reported that their participants achieved on the same task.

\xhdr{Dataset}
The final dataset $\mathcal{D}$ consists of (i) the criminal recidivism risk scores $\mathcal{S}^{i}_{j} \in \{1,2,..,5\}$ provided by our 20 respondents $i$ for 200 defendants $j$, as well as (ii) the COMPAS tool risk scores $\mathcal{S}^{C}_{j} \in \{1,2,..,10\}$ for 7214 defendants from the ProPublica dataset.
The dataset can be found at \url{https://github.com/hansen7/LearnFairMetric_Empirical}.

\subsection{Distance Metric Learning}
\label{sec:methodology}
In our experiments, we evaluate the performance of several different Mahalanobis metric learning approaches on our criminal recidivism dataset $\mathcal{D}$. 
To cover a broad range of learning methods, we considered one from each of the three learning paradigms discussed by \citep{bellet2013survey}: (i) fully supervised: Large Margin Nearest Neighbor (\emph{LMNN}, \citep{weinberger2006distance}); (ii) weakly supervised: Mahalanobis Metric for Clustering (\emph{MMC},  \citep{xing2003distance}); and (iii) semi-supervised: Least Squared-residual Metric Learning (\emph{LSML}, \citep{lsml}).

\xhdr{LMNN} \citep{weinberger2006distance}
The fully supervised \emph{LMNN} method can be directly applied on the labeled data provided by the ProPublica dataset and our respondents. 
It attempts to minimize the distance between training instances and neighbors of same class, while keeping instances of other classes out of the neighborhood. 
During the implementation, we consider that instances with the same rating in our dataset belong to the same class. In other words, in this approach, even though our data consists of Likert scale ratings, we treat these ratings as categorical values, thereby losing some information.

\xhdr{MMC} \citep{xing2003distance}
The weakly supervised \emph{MMC} method is designed to work in scenarios when rich labeled data is not readily available, and takes pairwise relative comparisons as inputs instead. 
The algorithm maximizes the sum of pairwise distances between dissimilar pairs while keeping that of similar pairs relatively small. The metric learned by \emph{MMC} can be constrained either in a diagonal form (weighted Euclidean) or as a full matrix.
We consider instances which have equal ratings in our dataset to be similar pairs, and the others to be dissimilar pairs. Again, as for \emph{LMNN}, this approach treats our Likert scale data as categorical values, disregarding some info.

\xhdr{LSML} \citep{lsml}
Unlike the \emph{LMNN} and \emph{MMC} methods, which can only utilize the 200 labeled instances, the semi-supervised metric learning algorithm \emph{LSML} allows the use of the remaining \textasciitilde 7000 unlabeled instances from the ProPublica dataset as well\footnote{We ran the algorithm both with and without using the 7000 unlabeled COMPAS data points. The results were qualitatively similar and we report the results of running the algorithm without using the unlabeled data.}.
It learns the metric from a set of triplet relative comparisons of the form "$a$ and $b$ are more similar than $a$ and $c$".
The triplet constraint set $\mathcal{C}$ is constructed as
$
\mathcal{C} = \{a, b, c|S_a \leq S_b + \sigma < S_c\}
$
, where $\sigma>0$ and $S_{a, b, c}$ are the recidivism scores from our dataset. 
Unlike \emph{LMNN} and \emph{MMC}, \emph{LSML} uses relative triplets, which allow us to capture more nuanced information available from our Likert scale judgments, such as "2 is closer to 3 than 5".

We implement an adapted version of the algorithm, by adding a trade-off coefficient $\alpha=0.01$ on the logdet regularization term.
This adaptation allowed us to reduce the weight of the regularizer, thereby increasing the weight for satisfying the relative triplet constraints.
As suggested by \citep{lsml}, we randomly subsampled the training inputs, instead of utilizing the full set $\mathcal{C}$ whose size is $\mathcal{O}(N^3)$, $N=140$. 

\xhdr{Procedure}
Each metric (\emph{LMNN}, \emph{MMC}, \emph{LSML}) is trained on 140 inputs and evaluated on 60 inputs. These 200 inputs were randomly selected. In our evaluation, we repeat this process 10 times and report the average results.

\section{Experiments}

\subsection{High-level Analysis of our Dataset}
As described in Section 3.1, we gathered data about a subset of 200 defendants from the ProPublica dataset in our survey. These 200 defendants have a similar distribution of demographic features as the full set of 7214 defendants from the ProPublica dataset. Below, we report a high-level analysis of our respondent's recidivism predictions (Q1), bail decisions (Q2), and decision confidence (Q3).

\xhdr{Recidivism Predictions and Bail Decisions} In Table 1, we investigate the relationship between the respondents' recidivism predictions and their bail decisions. In the first row, we see the fraction of defendants that were granted bail by our respondents. As expected, as the predicted risk of criminal recidivism increases, the propensity for granting bail decreases. However, in the second and third row, we find that different respondents have different thresholds for granting bail. For example, on average, our respondents granted bail to 18.9\% of defendants that they perceived as extremely likely to recidivate within 2 years. However, some respondents decided to grant bail to 76.5\% of such respondents, while others to 0\%.
\begin{table}[h]
\small

\renewcommand{\arraystretch}{1.4}
\centering
\begin{tabular}{c|ccccc}
\hhline{======}
 & & & \textbf{Recidivism Prediction} & & \\
 \textbf{Bail Rate} & \textbf{Extremely Unlikely} & \textbf{Unlikely} & \textbf{Neither} & \textbf{Likely} & \textbf{Extremely Likely} \\ \hline
Mean & 99.6\% & 96.6\% & 84.7\% & 43.1\% & 18.9\%\\
Max  & 100\% & 100\% & 100\% & 95.6\% & 76.5\%\\
Min  & 96.8\% & 77.9\% & 50.0\% & 0\% & 0\%\\
\hhline{======}
\end{tabular}
\caption{Relationship between recidivism predictions and bail decisions.}
\label{something}
\end{table}
\vspace{-0.5cm}

\xhdr{Bail Decisions and Decision Confidence} In Table 2, we explore the relationship between our respondents' bail decisions and their confidence in their decisions. The average accuracy of our respondents' bail decisions is 62.4\%. In the first column, we observe that our respondents tend to be slightly more accurate for decisions in which they have a higher confidence. However, in columns 2-4, we see that this relationship between confidence and accuracy varies between respondents -- some respondents have better calibrated confidence assessments than others. 
\begin{table}[h]
\small
\renewcommand{\arraystretch}{1.4}
\centering
\begin{tabular}{c|c|ccc|cc}
\hhline{=======}
\textbf{Accuracy} & \textbf{All Judges} & \textbf{Judge 1} & \textbf{Judge 10} & \textbf{Judge 18} & \textbf{Judge 8} & \textbf{Judge 17}\\ \hline
Overall & 0.624$\pm$0.027 & 0.635 & 0.580 & 0.580 & 0.536 & 0.640\\
High Confidence & \textbf{0.649$\pm$0.061} & \textbf{0.828} & \textbf{0.750} & \textbf{0.714} & 0.580 & 0.620\\
Low Confidence & 0.619$\pm$0.037 & 0.594 & 0.526 & 0.544 & \textbf{0.621} & \textbf{0.660}\\\hhline{=======}
\end{tabular}
\caption{Relationship between bail decisions and decision confidence.}
\end{table}
\vspace{-0.5cm}


\subsection{Evaluation}
In addition to the distance metric learning approaches discussed in Section \ref{sec:methodology}, we compare the three aforementioned metric learning approaches against two baselines: (i) the trivial baseline of the \emph{Euclidean} metric (i.e., the $\ell_2$ distance in the feature space), as well as (ii) the non-trivial, but naïve \emph{precision} matrix (i.e., the inverse of the covariance matrix), which is a standard approach for removing correlation between features\citep{mahalanobis1936generalized}. 
We evaluate the performance of the learned metrics with respect to the three loss functions defined below:

\xhdr{Relative Comparisons} This loss calculates the percentage of relative comparison triplets (constructed as described in the previous section) from the test set $\mathcal{C}_t$ which violate the constraints $d_\mathbf{M}(a, b)\leq d_\mathbf{M}(a, c)$. For the distance metric $\mathbf{M}$, it is similar to the loss defined by \citep{HofferA14}, where we assign equal weight to each instance:
\begin{equation}\label{the loss term}
    L(\mathbf{M}|\mathcal{C}_t) = \frac{H(d_\mathbf{M}(\mathbf{x}_a,\mathbf{x}_c)-d_\mathbf{M}(\mathbf{x}_a,\mathbf{x}_b))}{|\mathcal{C}_t|}\qquad\qquad \text{for  }\forall \{\mathbf{x}_a, \mathbf{x}_b, \mathbf{x}_c\} \in \mathcal{C}_t
\end{equation}
where $H(\cdot)$ is the Heaviside step function.

\xhdr{kNN L1} This loss calculates the $\ell_1$ divergence between the test instance ground truth label and the weighted rating of its neighbors, defined by the metric $\mathbf{M}$:
\begin{equation}
    L(\mathbf{M}|\mathcal{C}_t) = \sum_{\mathbf{x}_i\in\mathcal{C}_t}|\hat{y}_i - y_i| =\sum_{\mathbf{x}_i\in\mathcal{C}_t}|\sum_j w_{ij}y_j - y_j|
\end{equation}
where the normalised weight factor $w_{ij}$ is proportional to the inverse of the distance between $i$ and its k nearest neighbors $j$: $w_{ij}\propto 1/{d_\mathbf{M}(\mathbf{x}_i, \mathbf{x}_j)}$.

\xhdr{kNN L2} This loss is similar to kNN $\ell_1$ but instead calculates the $\ell_2$ distance. Compared to the $\ell_1$ norm, it introduces more penalty for large prediction errors.

\subsection{Metric Learning on Human Judgments}
For the collected human survey, we implemented \emph{LMNN, MMC, LSML} in Section 3.2 following the procedure described in Section 3.3.
The trade off coefficient $\alpha$ of the regularizer in our adapted \emph{LSML} is set to 0.01, and the number of neighbors for calculating \emph{kNN L1, kNN L2} is chosen to be five\footnote{We tried varying these parameters. They do not affect the high-level takeaways of our results.}.
The results are shown in Figure 1.

\begin{figure}[h]\label{fig:results}
\centering
\includegraphics[width=1\textwidth]{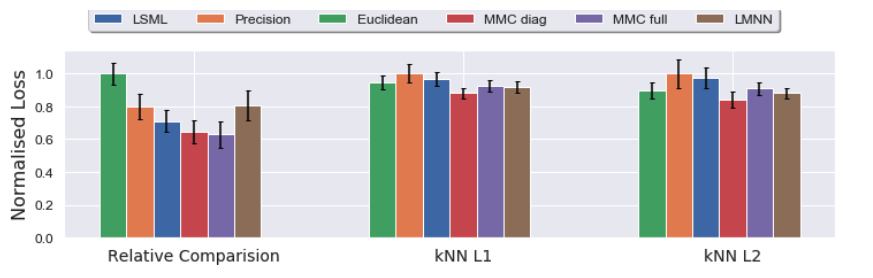}
\caption{Performance of the learned metrics, evaluated on our survey data.}
\end{figure}
\vspace{-0.3cm}

In Figure \ref{fig:results}, we observe that the learned metrics slightly outperform the Euclidean and Precision metrics with respect to the kNN L1 and L2 loss. For the triplet relative comparison loss, which incorporates the relative order between nearby ratings instead of treating them as categorical variables, the learned metrics have significantly outperformed the Euclidean and Precision metrics.

\subsection{Relative Comparisons on COMPAS}
In this section we evaluate the sensitivity of our adapted \emph{LSML} method to the hyperparameter $\sigma$, which controls the minimum required distance between inputs $b$ and $c$.
To this end, 
we compare the loss of the learned metric with the Euclidean metric on the COMPAS tool's predictions, from the ProPublica dataset, instead of the human judgments we gathered. Recall Section 3.2, which describes how we construct the sets of triplet constraints based on the choice of $\sigma$. The loss based on the relative comparisons is then calculated on the resulting $\mathcal{C}_t$. 


\begin{table}[h]
\centering
\small
\begin{tabular}{c|ccc}
\hhline{====}
$\sigma_t$ & \textbf{Euclidean} & \textbf{Ours($\sigma=0$)} & \textbf{Ours($\sigma=2$)}\\ \hline
0 & 0.40$\pm$0.037 & 0.39$\pm$ 0.041 & \textbf{0.38$\pm$ 0.035}\\
2 & 0.40$\pm$0.027 & 0.39$\pm$ 0.032 & \textbf{0.37$\pm$ 0.037}\\
4 & 0.35$\pm$0.031 & 0.31$\pm$ 0.045 & \textbf{0.30$\pm$ 0.034}\\
6 & 0.31$\pm$0.033 & \textbf{0.29$\pm$ 0.060} & \textbf{0.29$\pm$ 0.060}\\\hhline{====}
\end{tabular}
\caption{Loss of the metric learned using the adapted \emph{LSML} algorithm on the ProPublica dataset. Here $\sigma$ and $\sigma_t\geq 0$ represent the threshold values for constructing triplet constraints from the train and test set respectively.}
\end{table}
\vspace{-0.3cm}

Our learned metrics outperform the Euclidean distance by a large margin. As $\sigma_t$ increases, the loss decreases for all three metrics, since the difference between $b$ and $c$ increases.

\section{Discussion}
In this work, we conducted a user study, in which we gathered a set of human judgments about recidivism risk.
We initiated work to examine various methods for learning an individual fairness metric from human annotated data. 
Surprisingly, we observed similar performance across methods when considering predictive performance of ratings, though we saw differences when considering triplet consistency of unseen data. In future work, it would be interesting to understand this better and consider the interplay between the metric-learning methods and the consistency of human ratings.

\section*{Acknowledgements}
HW acknowledges support from Cambridge Trust CSC Scholarship. AW acknowledges support from the David MacKay Newton research fellowship at Darwin College, The Alan Turing Institute under EPSRC grant EP/N510129/1 \& TU/B/000074, and the Leverhulme Trust via the CFI. 
This research was supported in part by a European Research Council (ERC) Advanced Grant for the project "Foundations for Fair Social Computing", funded under the European Union's Horizon 2020 Framework Programme (grant agreement no. 789373).

\medskip
\small
\bibliographystyle{plainnat}
\bibliography{neurips_2019}

\end{document}